\documentclass[%
 reprint,
 amsmath,amssymb,
 aps,
]{revtex4-1}

\usepackage{graphicx}
\usepackage{subcaption}
\usepackage{dcolumn}
\usepackage{bm}
\usepackage{stmaryrd}
\usepackage{wrapfig}
\usepackage[mathlines]{lineno}

\begin{document}

\preprint{APS/123-QED}

\title{A self-consistent numerical approach to track particles \\ in FEL interaction with electromagnetic field modes}
\author{A. Fisher}%
\author{P. Musumeci}
\affiliation{%
 University of California at Los Angeles, Los Angeles, CA, 90066
}%
\author{S.B. Van der Geer}%
\affiliation{Pulsar Physics, Eindhoven, The Netherlands}

\date{\today}

\begin{abstract}
In this paper we present a novel approach to FEL simulations based on the decomposition of the electromagnetic field in a finite number of radiation modes. The evolution of each mode amplitude is simply determined by energy conservation. The code is developed as an expansion of the General Particle Tracer framework and adds important capabilities to the suite of well-established numerical simulations already available to the FEL community. The approach is not based on the period average approximation and can handle long-wavelength waveguide FELs as it is possible to include the dispersion effects of the boundaries. Futhermore, it correctly simulates lower charge systems where both transverse and longitudinal space charge forces play a significant role in the dynamics. For free-space FEL interactions, a source dependent expansion approximation can be used to limit the number of transverse modes required to model the field profile and speed up the calculation of the system's evolution. Three examples are studied in detail including a single pass FEL amplifier, the high efficiency TESSA266 scenario, and a THz waveguide FEL operating in the zero-slippage regime.
\end{abstract}

\maketitle

\section{Introduction}

Numerical simulations have played a significant role in the development of X-ray Free Electron lasers \cite{mcneil:xfel}. As the theory underlying the FEL \cite{huang:review,pellegrini:theory} only admits analytical solutions under strong approximations, accelerator physicists have over the years developed a well assorted suite of numerical approaches to better understand the details of the evolution of charged particles and electromagnetic fields in their interaction through magnetic undulators.

There are a large variety of FEL simulation codes and many good reviews on the subject have been given \cite{biedron:FELcomparison,reiche:felreview,giannessi:review}. These range from fast one dimensional models (Perseo \cite{giannessi:Perseo}, Perave \cite{emma:perave}) which help in quick design studies and can be used to explore time-dependent and non linear effects, to more complete 3D simulations (Ginger \cite{fawley:ginger}, Genesis 1.3 \cite{reiche:genesis}, Fast \cite{saldin:fast}, Puffin \cite{campbell:puffin}, Minerva \cite{freund:minerva}) which include transverse effects and can simulate wakefields and complex beam distributions with correlations between the phase spaces. Each code has been (at least initially) developed to solve a particular FEL problem, but it has often been the case that, by comparing and understanding the various assumptions in each model, insights on the various physical processes taking place in an FEL system have been gained.

Here we discuss yet another instance of a three dimensional FEL simulation based on the decomposition of the electromagnetic field in a discrete set of transverse and frequency modes. In this respect it is more similar to the family of frequency-based codes like Puffin or Minerva. The code is built as an expansion of the widely available General Particle Tracer code for charged particle simulations \cite{pulsarphysics}. In this sense, it can use a complete set of already built-in functions for beam transport and interface seamlessly with photoinjector \cite{bazarov:multivariate} and CSR calculations \cite{brynes:CSRGPT}. This choice also brings several important advantages. The calculation does not resort to period averaging and a full (simulated or even measured) undulator field map can be used to move the particles. The effects of the interaction at the undulator entrance and exit can therefore be correctly evaluated. Furthermore, space charge effects are naturally incorporated, including the transverse space charge effects that at low beam energy play a significant role in the beam transport and evolution.

The code can be used to simulate both free-space and waveguide propagating electromagnetic fields and can take into account the dispersive properties of the medium. For free-space there is some freedom in choosing the basis for the field expansion, making it possible to take advantage of the Source Dependent Expansion \cite{sprangle:sde, baxevanis:sde} algorithm to reduce the number of modes needed to accurately describe the field and significantly speed up the calculation.

The paper is organized as follows. We first review the modal expansion and the equations implemented in the simulation \cite{gover:RMP}. We then make three different application examples. The first one is just a simple seeded FEL amplifier in vacuum (analyzed both in helical and planar geometry). The second one applies to the study of the system in the strong non linear regime and refers to the simulation of the TESSA266 experiment \cite{TESSA266}. The final example is a waveguide THz FEL where the code is used to correctly simulate the zero-slippage amplification \cite{Curry:NJP}.

\section{\label{ModeExpansion} Mode Expansion}

In order to self-consistently simulate the interaction between radiation and electrons, we begin with the Maxwell wave equation for the complex field amplitude
\begin{align}
    \left( \nabla_\perp^2 +\frac{\partial^2}{\partial z^2}-\frac{1}{c^2}\frac{\partial^2}{\partial t^2}\right)  E(\vec x,z,t)  =\mu_0 \frac{\partial  \vec{\mathbf{J} }(\vec x,z,t)\cdot \hat{\mathbf{n}}^*}{\partial t}
    \label{Maxwell}
\end{align} 
where $\hat{\mathbf{n}}$ and $\vec{\mathbf{x}}$ denote the polarization vector and transverse coordinates, respectively. Defining $\hat{\mathbf{z}}$ as the direction of propagation, the polarization vector can be written in complex notation as $\hat{\mathbf{n}} =\hat{\mathbf{x}}$ or $\hat{\mathbf{n}}=(\hat{\mathbf{x}}\pm i \hat{\mathbf{ y}})/\sqrt{2}$ for linearly and circularly polarized light. The polarization vector formalism is particularly convenient to unify the description of the planar and helical geometries. The time-averaged Poynting vector (representing the wave intensity) can be written in both cases as $\epsilon_0 c |E(\vec x,z,t)|^2/2$.

If we write the scalar field amplitude in terms of its z-coordinate spatial Fourier transform
\begin{equation}
    \quad E(\vec x,z,t)=\frac{1}{2\pi}\int_{-\infty}^\infty \hat E(\vec x,k,t)e^{ikz-i\omega t}dk
    \label{spatial_ft},
\end{equation} 
the left hand side (LHS) of the equation can be rewritten as
\begin{multline}
    LHS=\frac{1}{2\pi}\int_{-\infty}^\infty\left(\nabla_\perp^2-k^2 + \omega^2/c^2+\frac{2i\omega}{c^2}\frac{\partial}{\partial t}\right) \\
    \hat  E(\vec x,k,t) e^{ikz-i\omega t}dk \label{LHSfinal}
    \end{multline}
where we factored out the harmonic time-dependence and have neglected the second derivative of the slowly varying field amplitude, i.e. $\partial^2 E(\vec x, k, t) / \partial t^2 \ll \omega^2 E(\vec x, k,t)$.

The current density on the RHS can be written in complex notation using the particle positions and velocities
\begin{align}
\vec{\mathbf{J}}(z,t)=\sum_j q_j \vec{\mathbf{v}}_j\delta(\vec x-\vec x_j(t))\delta(z-z_j(t)),
\label{current_density}
\end{align} 
where $\vec{\mathbf{v}}_j= \sqrt{2} K_{rms} c e^{-ik_u z_j}/\gamma_j \hat{\mathbf{n}}$ represents the particle velocities in the undulator, $K_{rms}=eB_{rms}/mck_u$ is the root mean square (rms) undulator strength parameter,  $\lambda_u=2\pi/k_u$ is the undulator period, $e$ and $m$ are the charge and mass of an electron, and $\gamma_j$ is the relativistic factor. Note that in most simulations a macroparticle model is used where one simulation particle represents multiple actual electrons in the beam. In this case, the sum in Eq. \ref{current_density} will run over the macroparticle index.

Using nested Fourier transforms, we have
\begin{align}
    RHS=\frac{\mu_0}{2\pi}\frac{\partial}{\partial t} \left[ \iint_{-\infty}^\infty  \vec{\mathbf{J}}(\vec x, z',t) e^{-ikz'}dz' e^{ikz}dk\right]\cdot \hat{\mathbf{n}}^*.
\end{align}
The delta function allows easy integration over $z'$. The time derivative is straightforward using chain rule with $z_j(t)$ after noticing that $K$ and $\gamma$ have a very slow dependence on $z_j$ ($\frac{dK}{dz} \ll k_u$ and $\frac{d\gamma}{dz} \ll k_u$) and the transverse velocity is negligible.
\begin{multline}
    RHS=\frac{-i\mu_0}{2\pi}\int_{-\infty}^\infty \sum_j q_jc\beta_{z,j}(k_u+k) (\vec{\mathbf{v}}_j \cdot \hat{\mathbf{n}}^*) \\ \times \delta(\vec x-\vec x_j)e^{-ikz_j+ikz}dk
    \label{RHSfinal}
\end{multline}

Combining Eqns. \ref{LHSfinal} and \ref{RHSfinal}, we can then rewrite Eq. (\ref{Maxwell}) for the spatial frequency components of the field as  
\begin{equation}
    \left(\nabla_\perp^2-k^2+\frac{\omega^2}{c^2}+\frac{2i\omega}{c^2}\frac{\partial}{\partial t}\right)\hat E(\vec x,k,t) = S(\vec x,k,t) 
    \label{field_equation}
\end{equation}
where the source term is obtained by projecting the current density onto $\mathbf{\hat{n}}$ as
\begin{multline}
S(\vec x,k,t)=
 \\ \sum_j -i\mu_0 c q_j \beta_{z,j}(k_u+k) (\vec{\mathbf{v}}_j \cdot \hat{\mathbf{n}}^*)\delta(\vec x-\vec x_j)e^{-ikz_j+i\omega t}. 
    \label{source_FEL}
\end{multline}

Each spatial frequency component of the field can be further decomposed into an orthogonal mode basis labeled by index $m$ and normalized such that $\iint \Theta^*_m \Theta_n d\vec x=\delta_{mn} A_m$ where $\Theta_m(\vec x,k,t)$ is one of the complex mode solutions of the source-free wave equation (i.e. $S$ = 0 in Eq. (\ref{field_equation})) and $A_m$ is a normalization constant. 

Inserting $\tilde E( \vec x, k, t)=\sum_m a_m(t) \Theta_m( \vec x, k, t)$ into Eq. (\ref{field_equation}), we can multiply both sides of the equation by $\Theta_n^*$ and integrate over the transverse coordinates $\iint  d\vec x$ to obtain the mode amplitude excitation equation \cite{gover:RMP}
\begin{align}
\label{excitation_equation}
    \dot a_m =-\sum_j\frac{q_j}{2\epsilon_0 A_m} \left[ \frac{c\beta_{z,j} (k_u+k)}{\omega}\right] (\vec{\mathbf{v}}_j \cdot \hat{\mathbf{n}}^*) \Theta_{m,j}^*  e^{-ikz_j+i\omega t}
\end{align}
where $\Theta_{m,j}$ means evaluating the m-th mode at the jth particle position. As we sum over the particles, only the spatial frequencies that are nearly resonant with the particle speeds ($\beta_{z_j}=\beta_{ph}=\omega/c(k+k_u)$) will contribute to a net energy exchange with the field so that the bracketed term can be approximated as 1.

This mode excitation equation can also be independently derived from (and is fully consistent with) energy conservation. To see this, we write the energy of the system $W$ using the spatial frequency Fourier transform of the electric field as
\begin{align}
    W=\frac{1}{2}\frac{\epsilon_0}{2\pi}\int \sum_m |a_m|^2A_m dk.
\label{GPT_energy}
\end{align}

After differentiating, we find
\begin{align}
    \frac{dW}{dt}=\int \sum_m \frac{a^*_m}{2\pi} \left[ \dot a_m \frac{\epsilon_0 A_m}{2}\right] dk + \textit{c.c.}
\end{align}

The rate of change in the electromagnetic energy is the negative of the work done on the particles,
\begin{align}
    \nonumber
    & \sum_j \vec{\mathbf{F_j}} \cdot \vec{\mathbf{v_j}}= -\sum_j q_j  \Re (E(\vec{x},z,t) \hat{\mathbf{n}}) \cdot \Re (\vec{\mathbf{v}}_j) \\ 
    &= \sum_j \frac{q_j}{4}\left(E^*(\vec{x},z,t) \hat{\mathbf{n}}^*\cdot\vec{\mathbf{v}}_j \right) + \textit{c.c.} \\
   & = \int \sum_m \frac{a^*_m}{2\pi} \left[\sum_j\frac{q_j}{4}\Theta_{m,j}^* e^{-ikz_j+i\omega t} (\hat{\mathbf{n}}^*\cdot \vec{\mathbf{v}}_j )\right] dk + \textit{c.c.}
\end{align}
where terms that do not satisfy the resonant condition average to zero in the particle sum. Equating the coefficients of $a^*_m$ leads to
\begin{align}
    \dot a_m=-\sum_j \frac{q_j}{2\epsilon_0 A_m} ( \vec{\mathbf{v}}_j\cdot \hat{\mathbf{n}}^*) \Theta_{m,j}^*e^{-ikz_j+i\omega t}
\end{align}
which matches our previous calculation in \eqref{excitation_equation}. In other words, the evolution of the amplitude of each electromagnetic mode in the system can be simply calculated by adding the energy changes induced by that mode on the particles. 

\subsection{GPT Numerical Implementation}
In order to extend the capabilities of GPT to self-consistently calculate the interaction with the radiation modes in the undulator, we based our development on the built-in function that computes the interaction with the modes of a gaussian optical resonator \cite{GPT_FEL}.

In the numerical model, the continuous integral of \eqref{spatial_ft} is approximated using a discrete basis of spatial frequency modes 
\begin{multline}
    \vec{\mathbf{E}}(\vec x,z,t) =\sum_q (u_q+iv_q) \Theta_q(\vec x, k, t) e^{ik_q z-i\omega_q t} \hat{\mathbf{n}}
\label{GPTfield}
\end{multline}
where the sum over index $q$ includes both spatial frequencies and transverse modes. With respect to the previous section, $u_q$ and $v_q$ now represent the actual electric field amplitudes and have absorbed the user-defined mode separation interval $\Delta k$ and the $1/2\pi$ from the Fourier transform. Consequently, the source term in Eq. \eqref{field_equation} also gains an additional factor of $\Delta k/2\pi$. 

In the input file, the user can specify the number of modes and the spatial frequency interval for the simulation. That choice of interval and associated spectral resolution should be taken judiciously to include the resonant frequency of the system and to correctly simulate the radiation bandwidth. Since the latter depends on various factors including the gain parameter, the length of the undulator, and the electron bunch length, it is always advisable to check the results for consistency and convergence as the number of modes and their separation is varied. 

The choice of the spatial frequency interval defines the distance in the z-dimension $L = 2\pi / \Delta k$ over which periodic boundary conditions are applied for the field. The frequencies $\omega_q$ are determined from the longitudinal wavenumbers using the mode dispersion relation given by $\omega_q = c k_q$ in free space or $\omega_q^2 = (k_{mn}^2 + k_q^2) c^2$ in a waveguide.

Writing the complex mode amplitude as $\Theta_q=T_q e^{i \psi_q}$, we can then express the $x$ and $y$ component of the electromagnetic field at time $t$ at the particle locations as
\begin{multline}
    E_x(\vec x_j , z_j,t)=\sum_q T_q\left( u_q\cos\phi_q-v_q\sin\phi_q\right) |\hat{\mathbf{n}} \cdot \vec{\mathbf{x}}|\\
    E_y(\vec x_j , z_j,t)=-\sum_q T_q\left( u_q\sin\phi_q+v_q\cos\phi_q\right) |\hat{\mathbf{n}} \cdot \vec{\mathbf{y}}| \\
    \vec{\mathbf{B}}=\frac{1}{\omega_q}\hat k_q \times \vec{\mathbf{E}} \\
\end{multline}
where $\phi_q = k_q z_j - \omega_q t + \psi_q$. 

From these fields, the electromagnetic forces acting on the particles are computed at each time step. Particle velocities and positions are then used to self-consistently calculate the evolution of the amplitudes of each mode ($u_q$ and $v_q$) according to Eq. \ref{excitation_equation}.

It is also possible to run the code in single frequency mode. In this case,  the field is assumed to be perfectly periodic, with only one spatial frequency term in Eq. \ref{GPTfield} and the time-averaged sum (now only running over the transverse modes) $\frac{\epsilon_0\pi}{\Delta k} \sum_q (u_q^2+v_q^2) A_q$ corresponds to the total radiation field power.

\subsection{Curved Parallel Plate Waveguide}
The geometry of the interaction to be simulated determines the choice of the mode basis $\Theta_m$ and the associated dispersion relation. An important application of our new code is the study of the evolution of an FEL system in a waveguide. The dispersive properties in the waveguide can not be easily modeled in conventional FEL codes which adopt a time-dependent (slice) model for the description of the radiation. For this case we can expand the field in the complete set of orthonormal modes for the particular waveguide cross-section under study. Here we focus on the TE modes of a curved parallel plate waveguide \cite{Curry:NJP} where the fields can be written in terms of the longitudinal component of the magnetic field $H_z$.

\begin{figure}[ht]
\includegraphics[width=\linewidth]{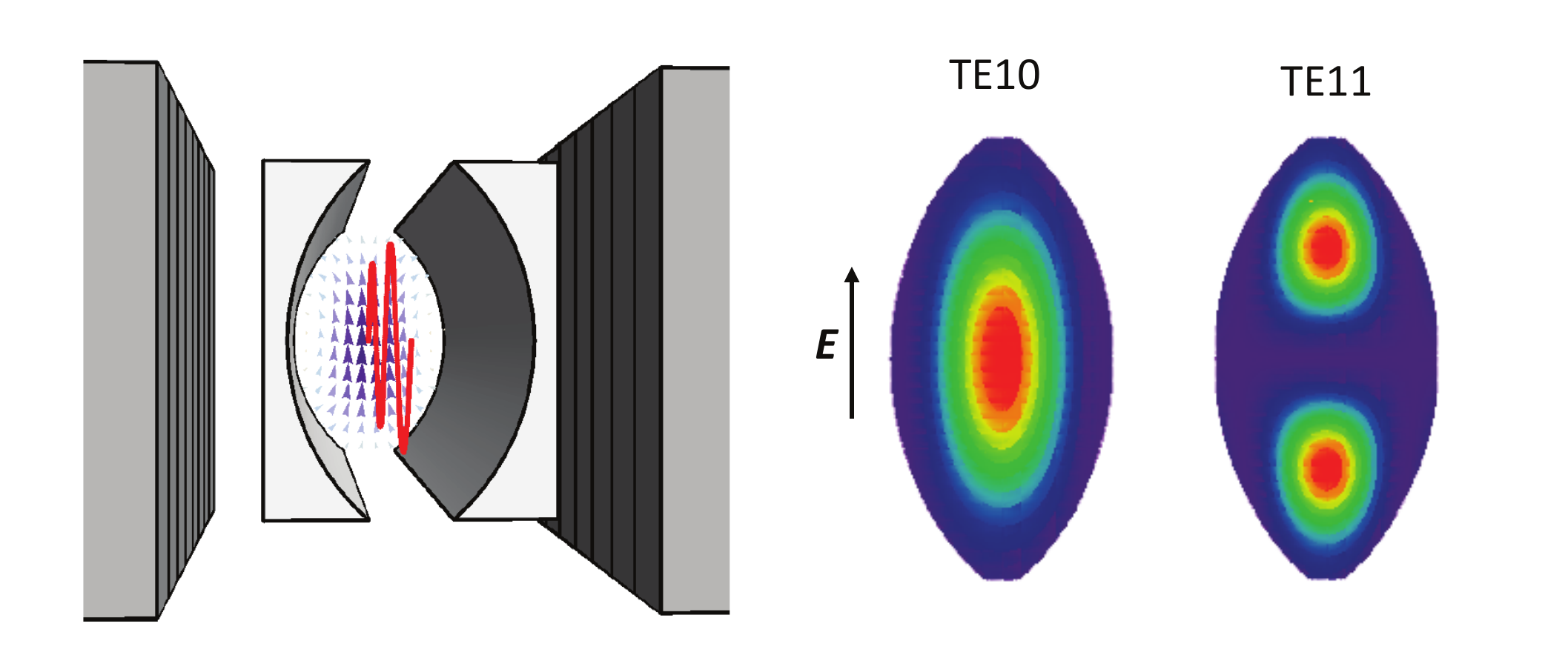}
\caption{TE10 and TE11 y-component of the electric field for a curved parallel plate waveguide. The TE10 mode is the one that has the largest FEL coupling to extract energy from a relativistic electron beam.}
\label{cppwg_modes}
\end{figure}

The transverse wavenumber $k_{mn}$ for the modes in the waveguide is written as 
\begin{equation}
k_{mn} = \frac{1}{b} \left( n\pi+(2m+1)\tan^{-1}\frac{b}{\sqrt{2Rb-b^2}} \right)
\end{equation}
where $b$ is the separation and $R$ is the radius of curvature of the waveguide. The confocal case (i.e $R = b$) minimizes diffraction losses and is typically employed in practice \cite{nakahara:cppwg}, but in the numerical model these parameters can be chosen by the user separately. 

The dispersion relation is then expressed as 
\begin{equation}
k_z(\omega) = \sqrt{\frac{\omega^2}{c^2}-k_{mn}^2}.
\end{equation}

Analytical expressions for the field  $\mathbf{\mathcal{E}}_{m,n}(\mathbf{r_\bot})$ in the guide can be found in \cite{nakahara:cppwg} and \cite{snively:thesis}. The longitudinal field $\Phi_{mn}$, corresponding to $H_z$ for TE modes and $E_z$ for TM modes can be written in terms of Hermite polynomials $He_m$ as 
\begin{multline}
    \Phi_{mn} = \frac{e^-\frac{\beta^2_{mn}x^2}{\alpha_{mn}(y)}}{\sqrt[4]{\alpha_{mn}(y)}}
    He_m\left(\frac{2\beta_{mn}x}{\sqrt{\alpha_{mn}(y)}}\right) e^{\pm ik_zz}\\
    \begin{bmatrix}\cos \\ \sin \end{bmatrix}
    \left[ k_{mn}y+\frac{2\beta^4_{mn}yx^2}{k_{mn}\alpha_{mn}(y)}-(m+\frac{1}{2})\arctan{\frac{2\beta^2_{mn}y}{k_{mn}}} \right] 
\end{multline}
where
\begin{equation}
\begin{split}
    \alpha_{mn}(y) = 1+4\frac{\beta_{mn}^4y^2}{k_{mn}^2}\\
    \beta_{mn}=\sqrt{\frac{k_{mn}}{\sqrt{2Rb-b^2}}}.
\end{split}
\end{equation}

The transverse field components are then calculated as
\begin{align}
    E_{(x,y)} = \frac{-i}{k_{mn}^2} \left( k_z \frac{\partial E_z}{\partial (x,y)} \pm \omega \mu \frac{\partial H_z}{\partial (y,x)} \right).
\end{align}

The effective mode area
\begin{equation}
\label{mode_normalization}
A_{mn} = \frac{\int |\mathcal{E}_{mn}(r_\bot)|^2 dr_\bot}{|E_{peak}|^2}
\end{equation}
is hard-coded in the software. 

\subsection{Free space propagation Source Dependent Expansion}
Another important case is where the waveguide boundaries are removed or very far away so that one can use free-space modes to describe the radiation field. Either Laguerre-Gaussian or Hermite-Gaussian modes can be used depending on the symmetry of the problem. Assuming azimuthal symmetry (i.e. $r^2 = |\vec{x}|^2$), we start by writing the complex scalar field amplitude as a sum of different spatial frequency Laguerre-Gaussian modes, 
\begin{align}
    E(\vec x,z,t)=\sum_{n,m}a_{n,m}(t) \Theta_{n.m}(r,t) e^{ik_n z-i\omega_n t}
\label{lg_expansion}
\end{align}
where we explicitly show that the sum index runs over the different spatial frequencies (n) and the transverse mode numbers (m). The modal basis for the field expansion can be written as  
\begin{multline}
\Theta_{n,m}(r,t)=\frac{1}{\sqrt{1+\alpha_n(t)^2}}L_m\left(\frac{2r^2}{w_n(t)^2}\right)e^{-r^2/w_n(t)^2} \\ \times e^{i\alpha_n(t)r^2/w_n(t)^2-i(2m+1)\psi_n(t)}
\end{multline}
where $L_m$ is the Laguerre polynomial of order $m$, $w_n$ and $\alpha_n$ indicate the waist size and the curvature of the phase fronts for the mode having spatial frequency $k_n$, and $\psi_n(t) = \arctan{\alpha_n(t)}$. In the case that no electron beam is present and the radiation is freely diffracting, $w_n(t)=w_{0,n}\sqrt{1+c^2t^2/z_{r,n}^2}$ and $\alpha_n(t)=ct/z_{r,n}$ with the implicit frequency dependence in $z_{r,n} = k_n w_{0,n}^2/2$, the Rayleigh range of the nth-mode. The mode area normalization constants are 
\begin{equation}
    A_{n,m} = \pi w_{0,n}^2 / 2.
\end{equation}

The effectiveness of the Laguerre-Gaussian mode expansion depends critically on the choice of the waist size and location, and in the absence of any prior knowledge or extra information, the simulation should include a large number of transverse modes in order to accurately model the radiation field. 

In many cases, as for example when the FEL is seeded with an external laser and the radiation transverse profile is mainly dominated by one or a few modes, it is a good approximation to truncate the sum to only include a small number of terms. To further minimize this number (and proportionally speed up the computational time), it is possible to take advantage of the source dependent expansion originally developed for the FEL framework by Sprangle et al. \cite{sprangle:sde} where the waist size and location of the expansion are adjusted along the interaction. 

Following the original work in \cite{sprangle:sde} (recently revisited by Baxevanis et al. \cite{baxevanis:sde}), after plugging Eq. \ref{lg_expansion} into the inhomogeneous wave equation, we obtain a coupled system of differential equations for the mode amplitudes in terms of the projections of the source term onto the mode basis. 
\begin{equation}
F_{m,n} =\frac{c^2}{\omega_n \pi w_{0,n}^2}\int S(r)\Theta_m^*(r) d \vec{x} \nonumber
\end{equation}
Using the definition of $S$ from \eqref{source_FEL}, it is possible to write the source projection moments $F_{m,n}$ in terms of sums over the particle (or macroparticle) coordinates.

We can then solve for how $w_n$ and $\alpha_n$ should vary in order to truncate the system at the desired order. For example, neglecting all $m\geq1$ we get 
\begin{align}
    &\frac{\partial u_n}{\partial t}=G_n\left( \alpha_n u_n-v_n\right)+(u_nB_{I,n}+v_nB_{R,n}) \nonumber \\ 
    &\quad \quad \quad + F_{0I,n} \nonumber\\
    &\frac{\partial v_n}{\partial t}=G_n\left( u_n+\alpha_n v_n\right)+(v_nB_{I,n}-u_nB_{R,n}) \nonumber\\
    &\quad \quad \quad - F_{0R,n} \nonumber \\
    &\frac{\partial \alpha_n}{\partial t}=\frac{2(1+\alpha_n^2)c^2}{\omega w_{n}^2}+2B_{R,n}-2\alpha_n B_{I,n} \nonumber \\
    &\frac{\partial w_{n}}{\partial t}=\frac{2c^2\alpha_n}{\omega_n w_{n}}-w_{n} B_{I,n}
    \label{SDE_equations}
\end{align}
where $G_n=\frac{2}{1+\alpha_n^2}(B_{R,n}-\alpha_n B_{I,n})$. $B_n$ represents the correction to the mode waist and radius induced by the source and can be written as  
\begin{align}
B_n&=F_{1n}e^{-2i\psi_n}/a_n. \label{Bfor1mode}
\end{align}

A closer inspection to Eq. \ref{SDE_equations} c and d indicates that $c/|B_n|$ is a distance which sets the scale for the variation of the mode radius. In multi-frequency simulations, the modes with small initial amplitudes cause the magnitude of $B_n$ to diverge. This is taken care of by setting a user-defined input parameter $L_{thresh}$ which limits the spot size variation along the interaction by setting $B_n=0$ whenever $c/|B_n|<L_{thresh}$. 

The equations for radiation evolution are then self-consistently solved with the GPT equations of motion for the macroparticles.


The general equations for complex mode evolution and $B_n$ with $M$ spatial modes are
\begin{align}
    \dot a_{n,m}& = \left[ B_{I,n}+\alpha_n G_n + i(2m+1)(G_n - B_{R,n}) \right] a_{n,m} \nonumber\\
   & + imB_n e^{2i\psi_n}a_{n,m-1} \nonumber\\
   & +i(m+1)B_n^* e^{-2i\psi_n} a_{n,m+1} - i F_{m,n} \nonumber\\
   a_{n,m\geq M}&=0 \implies B_n =\frac{F_{M,n}e^{-2i\psi_n}}{M a_{n,M-1}}.
   \label{modeevolution_multimode}
\end{align}
Higher order modes with small initial amplitudes are initially considered perturbations to the gaussian mode such that \eqref{Bfor1mode} still holds. Once the approximation $|a_1|/|a_0| \ll 1$ breaks down ($\approx .01$), the correct definition of $B_n$ from \eqref{modeevolution_multimode} can be used without divergence or significant numerical noise. In practice, errors from the perturbative approximation are negligible since it is accurate far into the linear regime.

\subsection{Quiet start}
In multifrequency simulations where many longitudinal wavenumbers and corresponding frequencies are used to simulate the field along a finite length bunch, it is critical to pay attention to the details associated with loading the particle coordinates in the simulation. Because it is common to have a much smaller number of macroparticles than real number of electrons, the noise in the bunching source term can be unacceptably high, causing unphysical growth of the field along the undulator.

\begin{figure*}[ht]
\includegraphics[width=\linewidth]{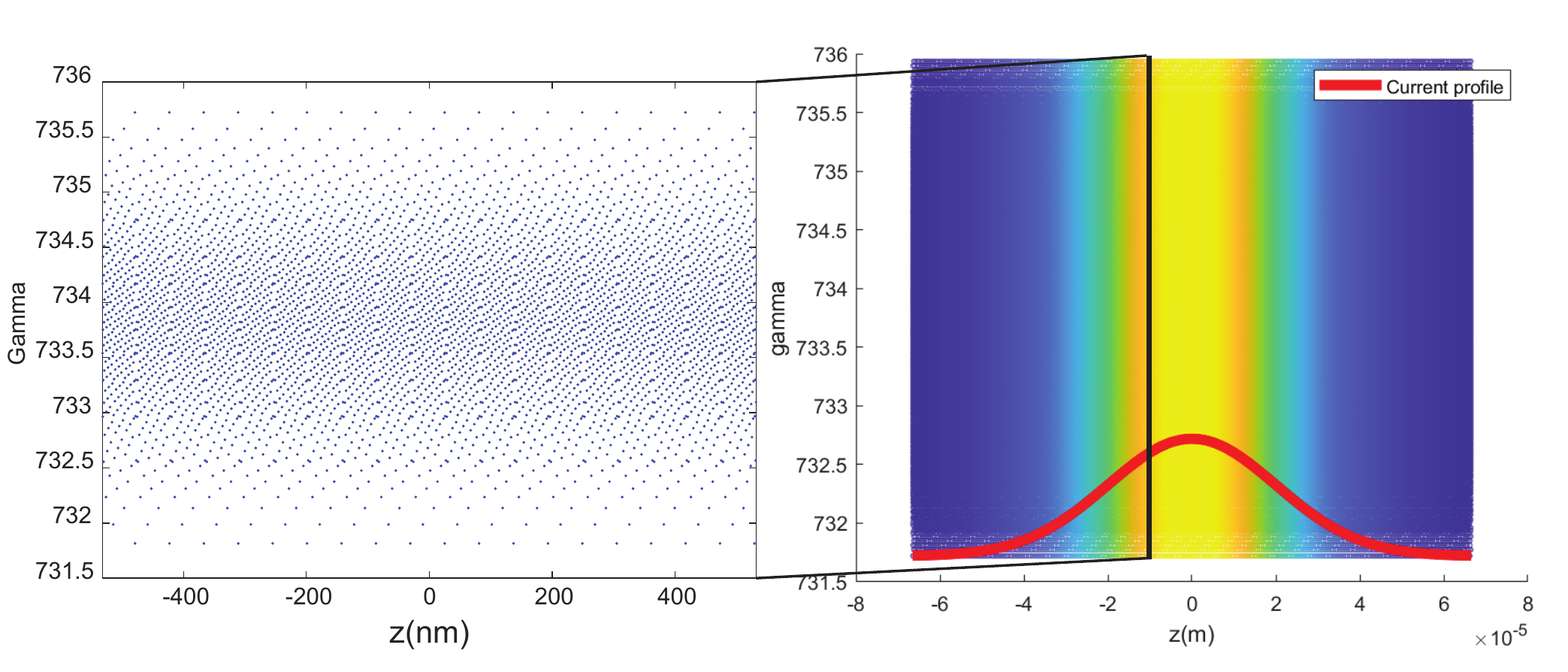}
\caption{left) Longitudinal phase space distribution with quiet loading for time-independent (i.e. single frequency) simulation. right) Longitudinal phase space distribution for multifrequency simulation. Particles are color coded by their charge weight. The projection onto the z-axis shows the Gaussian current profile.}
\label{quiet_start}
\end{figure*}

This problem is common and well discussed in the vast literature of simulations for FELs \cite{freund:shotnoise,mcneil:shotnoise}. While there are a number of possible solutions, our situation is slightly complicated as we need to ensure that the intrinsic bunching is and remains very small for all of the discrete frequencies in the simulation. This first requires equally distributing particles in the z-coordinate over a length $L = 2\pi/\Delta k$. For example in Fig. \ref{quiet_start} we show the input phase space when the simulation spans a bandwidth of 3 $\%$ around the central wavelength of 266nm. In this case, the beam longitudinal profile (a gaussian with rms bunch length 30 $\mu$m) is initialized by assigning a different charge weight to each macroparticle. When shot-noise effects are desired, each macroparticle's position is shifted by a small $dz$ according to well described algorithms \cite{fawley:shotnoise,penman:shotnoise} to achieve the correct statistics.

In addition, it is important to make sure that the noise from other coordinates would not contribute to a growth of the bunching as the beam propagates in the absence of an interaction. This is taken care of by mirroring the energy, transverse coordinates, and momenta over a large number of 5D phase space bins. The number of bins (typically larger than 32) should be chosen such that bunching in the absense of an interation remains small for all the discrete frequencies included in the simulations.

\section{Examples}
We limit this discussion to three examples that highlight the main features of our approach, even though it is expected that the new code can be successfully applied to a variety of other situations. The first case considered is a classical single-pass FEL seeded amplifier which will enable a quantitative comparison with the semi-analytical M. Xie formulas \cite{xie:formulas} as well as with a traditional period-average code like Genesis for both planar and helical geometries. The second example is relevant to the TESSA266 experiment being planned at the LEA beamline at the APS linac in Argonne National Laboratory aiming at very high conversion efficiency at 266 nm \cite{TESSA266}. This case serves to illustrate the capability of using a 3D magnetic field map for a fairly complicated segmented tapered undulator. The code compares well with a traditional FEL code like Genesis, even deep in the non-linear regime. The details of the beam transport (injection, entrance and exit sections and especially undulator break sections) can only be included in Genesis by using a linear beam transport approximation. GPT follows the evolution of the beam distribution along the beamline using field maps for all the magnetic elements (undulators, quadrupoles and phase shifter dipoles) and calculates energy exchange using the self-consistent interaction with the free-space modes. The results allow us to quantitatively include the effects of the entrance and exit sections (which add an effective 0.5 periods of interaction on each side of the undulator) and the trajectories after the prebuncher and in between the undulators. 

The final example is a waveguide THz FEL where GPT-FEL is used to correctly simulate the zero-slippage amplification. In this configuration, the strong dispersive properties of the guide affect the interaction which takes place in the zero-slippage regime. This scenario highlights a unique capability of our code which would be particularly challenging to simulate with traditional FEL codes.

\subsection{FEL amplifier}

\begin{figure*}[htb]
\centering
\begin{subfigure}[b]{0.47\linewidth}
\includegraphics[width=\linewidth]{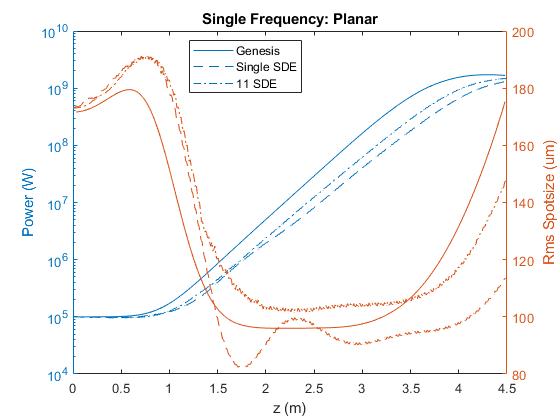}
\end{subfigure}
\begin{subfigure}[b]{0.47\linewidth}
\includegraphics[width=\linewidth]{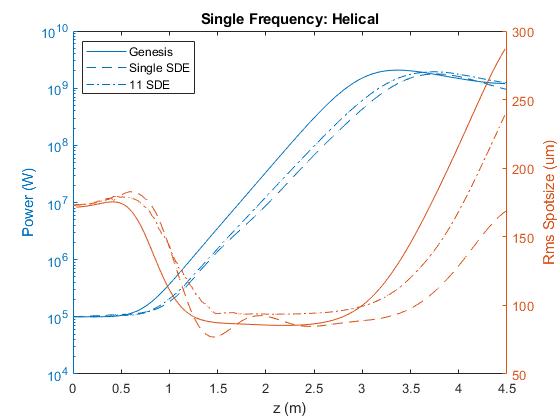}
\end{subfigure}
\caption{A comparison of GPTFEL running with SDE versus Genesis 1.3. a) The predicted gain length for the planar amplifier is 0.287 m. Simulating with SDE and a single spatial mode overshoots by 16\%. Running with 11 SDE spatial modes reduces the error to 5.9\%. b) The predicted gain length for the helical amplifier is 0.224 m. Simulating 1 and 11 SDE modes leads to errors of 15\% and 8.2\%, respectively.}
\label{fig:TI}
\end{figure*}

The parameters for this example are reported in Table \ref{tab:tessaparam} and somewhat arbitrarily chosen to be similar to an un-tapered version of the TESSA266 experiment discussed below. The main differences are that a 200 period long undulator (with no break-section) is used for this example and the input seed power is lowered to 10 kW. An analytical model for the undulator magnetic field is used. The beam is transversely matched to the undulator natural focusing (equally distributed in the horizontal and vertical plane) so that its rms spot size remains nearly constant along the interaction. The main goal of this example is to benchmark GPTFEL against the fitting formulas for the 3D gain length of an untapered FEL amplifier and compare with a conventional FEL code like Genesis. We also used this example to evaluate the performance of the single mode SDE approximation versus a simulation with $n = 11$ azimuthally symmetric Laguerre Gaussian SDE modes to decompose the electromagnetic field. GPTFEL took 1.5 minute to simulate 76800 particles on an 8 processor for the single SDE mode and 5 minutes for 11 SDE modes.

\begin{figure*}[htb]
\centering
\begin{subfigure}[b]{0.47\linewidth}
\includegraphics[width=\linewidth]{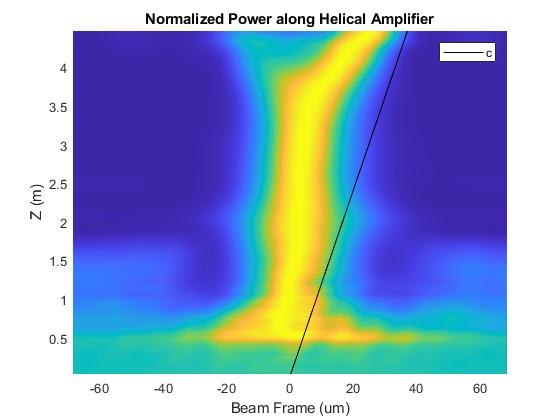}
\end{subfigure}
\begin{subfigure}[b]{0.47\linewidth}
\includegraphics[width=\linewidth]{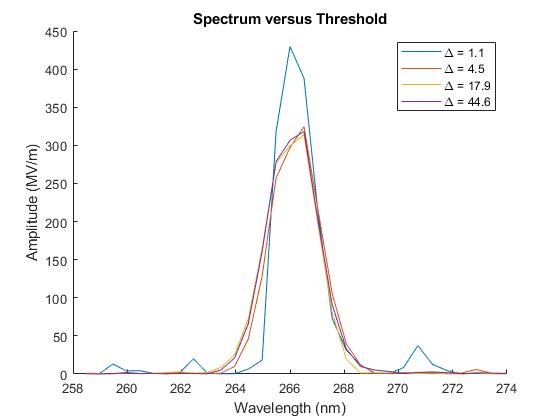}
\end{subfigure}
\caption{GPTFEL results for 31 spatial frequencies, each with a single gaussian transverse mode. a) Waterfall plot of normalized power. b) Spectrum at P=0.1 GW for different thresholds on SDE interaction. $\Delta$ is the ratio of $L_{thresh}$ to the theoretical gain length. Numerical errors occur when $\Delta\lessapprox 1$ because noise in the small amplitude, higher order modes quickly excite significant changes in the mode parameters. This suggests $L_{thresh}$ should be an order of magnitude larger than the theoretical gain length for convergent results. }
\label{fig:TD}
\end{figure*}

\begin{table}[htb]
\caption{\label{tab:tessaparam}Parameters for the 266 nm FEL amplifier simulation.}
\begin{tabular}{c c}
    \hline
    \multicolumn{2}{c}{Electron Beam} \\
    \hline
    Energy     & 375.5 MeV  \\
    Energy Spread     & 0.1 \% \\
    RMS Bunch length & 20 $\mu$m \\
    $\epsilon_{n,x},\epsilon_{n,y}$ & 2 mm$\cdot$mrad \\
    $I_{peak}$ & 1 kA \\
    $\sigma_x,\sigma_y$ & 72.5 $\mu$m \\
    \hline
    \multicolumn{2}{c}{Radiation} \\
    \hline
    $\lambda_1$ & 266 nm \\
    Input Power & 10 kW \\
    Rayleigh Length & 1.41 m \\
    Waist location & 0 m\\
    \hline
    \multicolumn{2}{c}{Undulator} \\
    \hline
    $K_{rms}$ & 2.82  \\
    $\lambda_u $& 0.032 m \\
    \hline
    \end{tabular}
\end{table}

The time-independent, single frequency results for the planar and helical geometries are shown in Figure \ref{fig:TI} and compared with Genesis 1.3. When using multiple spatial modes, the gain lengths in the planar and helical case are in good agreement (within 10 $\%$) of the semi-analytical and numerical model predictions. The radiation spot sizes defined by $\sigma_r^2=\frac{1}{2}\frac{\int r^2|E|^2 d^2\mathbf{x}}{\int |E|^2 d^2\mathbf{x}}$ also closely follow the prediction. Note that while a single SDE mode is able to achieve qualitative results up to and near saturation, a larger number of spatial modes is required to correctly simulate the evolution of the radiation profile after saturation.

The multi-frequency simulation used an SDE gaussian mode for 31 spatial frequencies with a 6\% bandwidth to simulate 128,000 particles in 23 minutes. The user-defined parameter $L_{thresh}$  limits the spot size variation along the undulator. Figure \ref{fig:TD}a shows a waterfall plot in the electron beam frame normalized at each z position to display the relative velocity of the radiation wavepacket, which is close to the beam velocity in the exponential regime and becomes superluminal in the non linear regime \cite{yang:superluminal}. In Figure \ref{fig:TD}b, the spectrum just before saturation is shown as a function of $L_{thresh}$ normalized to the gain length. If an increased spectral resolution is required, computation time scales linearly with number of tracked modes. 


\subsection{TESSA266}

\begin{figure*}[htb]
\centering
\begin{subfigure}[b]{0.47\linewidth}
\includegraphics[width=\linewidth]{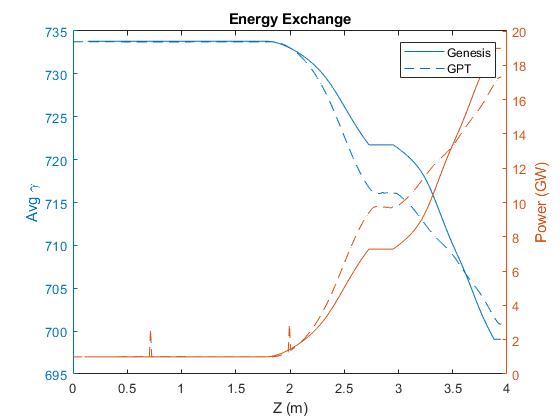}
\end{subfigure}
\begin{subfigure}[b]{0.47\linewidth}
\includegraphics[width=\linewidth]{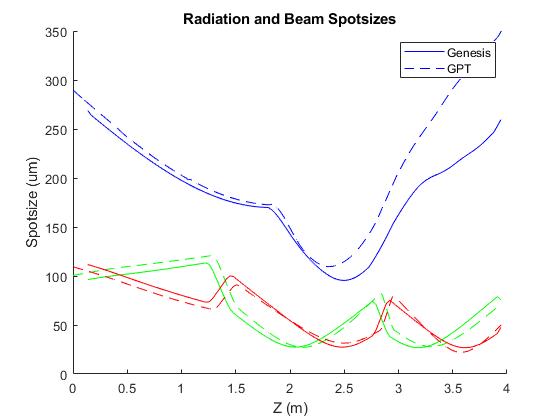}
\end{subfigure}
\caption{Energy exchange and spotsizes in the first two tapered undulators of the TESSA beamline.}
\label{fig:TESSAEnergyandSpotsize}
\end{figure*}

\begin{figure*}[htb]
\centering
\begin{subfigure}[b]{0.47\linewidth}
\includegraphics[width=\linewidth]{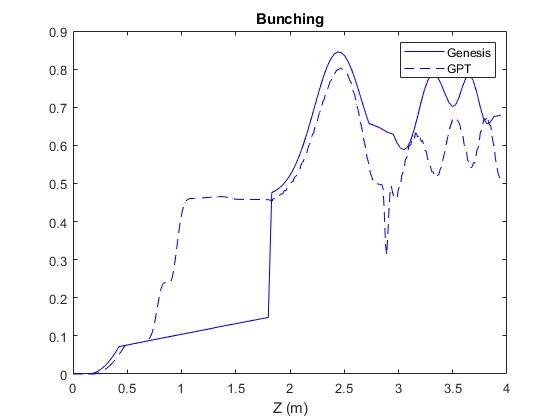}
\end{subfigure}
\begin{subfigure}[b]{0.47\linewidth}
\includegraphics[width=\linewidth]{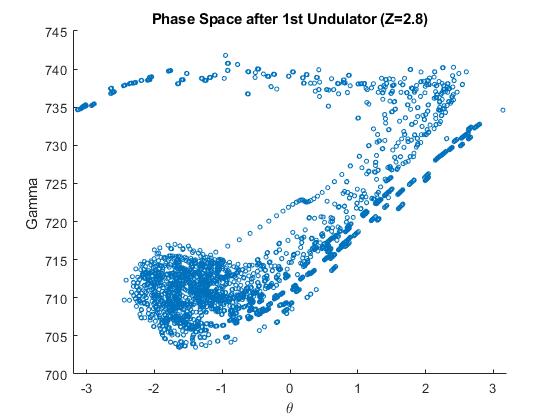}
\end{subfigure}
\caption{Bunching and Phase Space from the TESSA beamline.}
\label{fig:TESSABunchingPhaseSpace}
\end{figure*}

\begin{figure*}[ht]
\includegraphics[width=\linewidth]{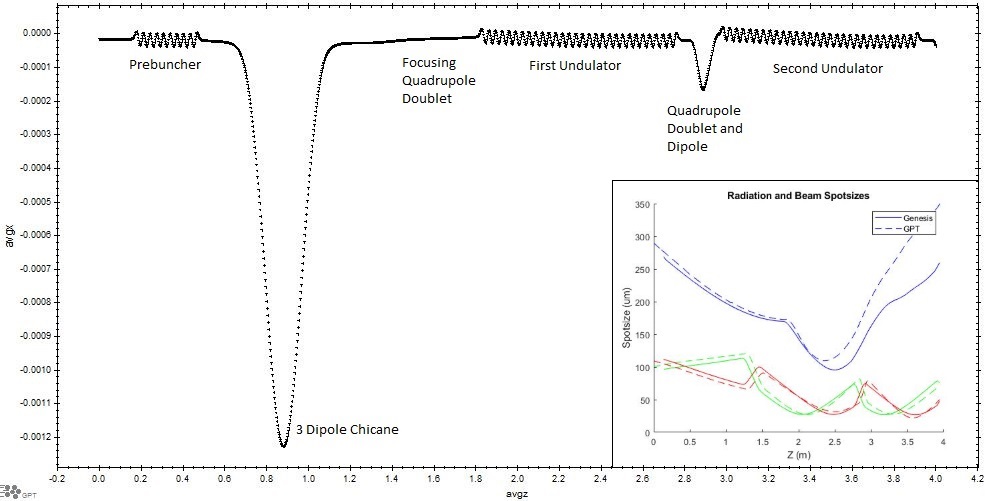}
\caption{TESSA Beamline}
\label{fig:TESSABeamline}
\end{figure*}

In this next example we take advantage of the GPT functions to track the electron beam in the fairly complex transport line of the TESSA 266 experiment. The beamline includes a short, 8 period undulator followed by a 3 dipole chicane to convert the imprinted energy modulation into microbunching. Quadrupole doublets match the beam transversely into the focusing channel of each $0.96$ meter, strongly tapered undulator section. A small dipole is placed between the second quadrupole doublet so that the three magnets can be used as a phase shifter between the undulator sections.

The GPT transport functions are used to set up the trajectory and the beam optics prior of turning on the seed and the FEL interaction module. Our time-independent simulation of the TESSA266 beamline includes 21 higher order spatial modes to ensure an accurate modeling of the radiation profile. A 1 GW peak power input radiation pulse is focused at the entrance of the tapered undulator to a waist of 0.3 mm. The simulation is compared with Genesis results, but it should be noted that GPTFEL uses full 3D magnetic field maps for the undulators as well as for the dipoles and quadrupoles in the system. The magnetic field in the chicane dipoles is fine tuned to maximize the bunching and simultaneously optimize the injection phase of the beamlets relative to the radiation phase at the entrance of the tapered undulator. In Genesis, both the R56 and phase shifts are applied post-facto to the beam distribution at the entrance of the tapered undulator, explaining the large difference in the bunching factor evolution in Fig. \ref{fig:TESSABunchingPhaseSpace}a. In practice, the phase shifter between the tapered undulator sections had to be re-optimized to account for the additional slippage incurred by the beam when passing in the entrance and exit section of the wigglers. This is accomplished by horizontally shifting the quadrupoles in opposite directions to steer the beam and tuning the magnetic field amplitude of the dipole to recover a straight trajectory while maximizing the energy exchange in the second undulator. 

\subsection{Zero slippage THz FEL}

A final example to showcase the capabilities of the new GPTFEL code is the simulation of a THz FEL operating in the zero-slippage regime \cite{snively:oe}. The size of the waveguide is chosen in order to match the group velocity of the radiation with the electron beam longitudinal velocity inside the undulator. This increases the bandwidth of the resonant interaction and extracts a significant amount of energy from very short electron beams.

\begin{figure}[h!]
\includegraphics[scale=0.45]{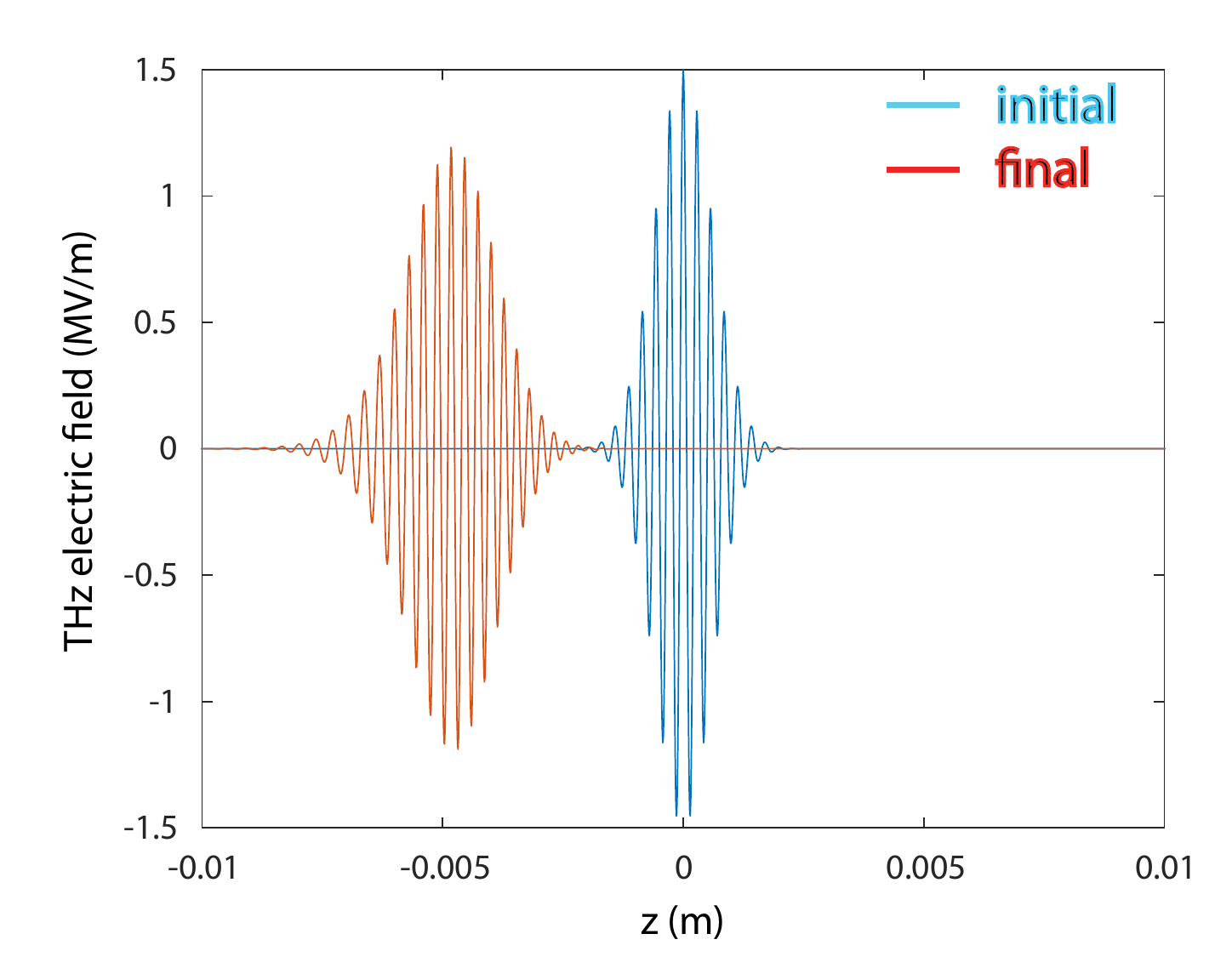}
  \caption{GPT time-dependent simulation temporal field profile at the entrance and at the exit of the 1 m long waveguide. The shift in the peak corresponds to the group velocity difference from the speed of light which is matched to the electron beam longitudinal velocity in the undulator in the zero-slippage regime. Helical geometry. Radiation spectrum and temporal profile of the pulse along the undulator. Final longitudinal phase space.}
  \label{cppwg_dispersion}
\end{figure}

GPTFEL correctly simulates the waveguide dispersive properties as shown in Fig. \ref{cppwg_dispersion} by plotting the electric field at the entrance and exit of the 1 meter long waveguide system in the absence of strong interaction (i.e. for very low charge beams). 

The parameters of this example are summarized in Table \ref{tab:THzparam}. We have chosen a planar undulator geometry with equally distributed focusing in the horizontal and vertical plane. In this case, the largest coupling is obtained with the TE10 mode profile of the curved parallel plate waveguide. The beam is initialized at the entrance of the simulation with a large bunching factor (0.5) while we set the amplitude of the initial input seed to zero.

\begin{figure*}[htb]
\includegraphics[width=\linewidth]{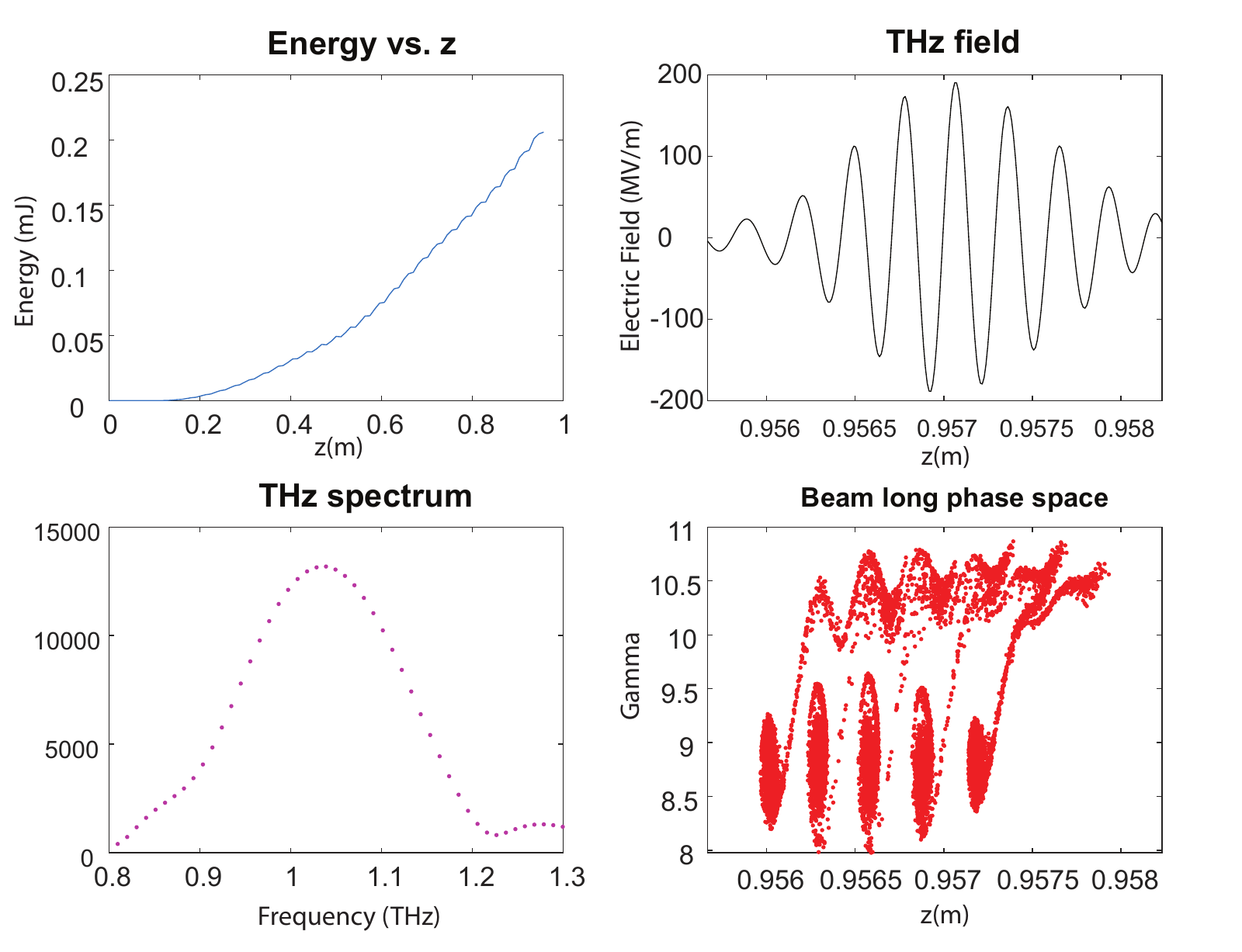}
\caption{a) THz pulse energy along the undulator. b) THz waveform at the undulator exit. c) THz spectrum. d) Longitudinal phase space of electron beam.}
\label{THz_simulation}
\end{figure*}

\begin{table}[htb]
\caption{\label{tab:THzparam}
Parameters for high efficiency THz amplifier.}
    \begin{tabular}{c c}
    \hline
    \multicolumn{2}{c}{Electron Beam} \\
    \hline
    Energy     & 10.2 MeV  \\
    Energy Spread     & 1.25 \% \\
    Bunch length & 2000 $\mu$m \\
    $I_{peak}$ & 60 A \\
    $\epsilon_{n,x},\epsilon_{n,y}$ & 5 mm$\cdot$mrad \\
    $\sigma_x,\sigma_y$ & 120 $\mu$m \\
    \hline
    \multicolumn{2}{c}{Undulator and waveguide} \\
    \hline
    $K_{rms}$ & 1.556  \\
    $\lambda_u $& 0.032 m \\
    $b$ & 1.9 mm \\
    $R$ & 1.9 mm \\
    \hline
    \end{tabular}
\end{table}

There are two main advantages of using the waveguide in this system. First, the waveguide maintains a constant radiation cross section along the interaction, avoiding diffraction effects. Second the waveguide's dispersive properties enable a zero slippage interaction. This large bandwidth interaction can drive the FEL with a much shorter beam because the slippage effects are effectively minimized and the radiation continues to interact and exchange energy with the particles even after a large number of periods. The simulation results are shown in Fig. \ref{THz_simulation} where the THz electric field waveform and the electron beam longitudinal phase space are shown to be temporally overlapping at the end of the undulator. Note that the system evolves in the non linear regime from the beginning as the electron beam enters the undulator with a very large bunching at the 1 THz resonant frequency induced by modulating the photocathode drive laser \cite{Musumeci:nlsco}. The undulator is linearly tapered starting from its half way point with a relative change in normalized vector potential $K$ of 30 $\%$/m to avoid saturation effects due to particles falling off the resonance curve. The efficiency of conversion is above 10 $\%$ in this example. 

\section{Conclusions and outlook}
A new approach for FEL simulations has been presented. The characteristic features are the decomposition of the field in a set of spatial and frequency modes and the integration with the GPT numerical integration engine which allows access and compatibility with a large number of beam transport designs and functions. There are a number of research opportunities which go beyond the scope of this paper but will be the subject of future studies, including a detailed study of the effects of the transverse space charge forces and an upgrade to include higher harmonic interactions. Parallelization of the code will allow much faster run times, increasing the number of macroparticles and modes that can be simulated. GPTFEL is not expected to replace traditional approaches to FEL numerical simulations, but is intended to be a research tool to explore the interaction of relativistic electrons and electromagnetic waves in undulator systems in regimes where the approximations of standard FEL codes are questionable. The application of GPTFEL to dispersive systems allows for exploration of novel interaction regimes like the tapered waveguide THz FEL.

\begin{acknowledgments}
The authors would like to thank Luca Giannessi and Avraham Gover for their helpful discussions. This work was supported by DOE grant No. DE-SC0009914. 
\end{acknowledgments}

\bibliographystyle{apsrev4-1}
\bibliography{GPTFEL}

\end{document}